\documentstyle[twocolumn,epsf,aps]{revtex}
\tighten
\begin{document}
\draft
\title{Lattice-gas model for alkali-metal fullerides:\\
face-centered-cubic structure}
\author{L\'aszl\'o Udvardi}
\address
{Quantum Theory Group, Institute of Physics, Technical
University of Budapest,\\
H-1111 Budapest, Budafoki \'ut 8, Hungary}
\author{Gy\"orgy Szab\'o}
\address
{Research Institute for Materials Science, H-1525 Budapest,
POB 49, Hungary}
\date{\today}
\address{
\centering{
\medskip \em
\begin{minipage}{15.4cm}
{}~~~A lattice-gas model is suggested for describing the ordering
phenomena in alkali-metal fullerides of face-centered-cubic
structure assuming that the electric charge of alkali ions
residing in either octahedral or tetrahedral sites is
completely screened by the first-neighbor C$_{60}$
molecules. This approximation allows us to derive an
effective ion-ion interaction. The
van der Waals interaction between the ion and C$_{60}$
molecule is characterized by introducing an additional site
energy at the tetrahedral sites. This model is investigated
by using a three-sublattice mean-field approximation and a
simple cluster-variation method. The analysis shows a large
variety of phase diagrams when changing the site energy
parameter.
\end{minipage}
}}
\maketitle
\narrowtext

\section{INTRODUCTION}
\label{sec:intro}

Since the discovery of superconductivity in several
alkali-metal fullerides \cite{hebard,ross,holczer} a large
variety
of A$_x$C$_{60}$ intercalation compounds has been prepared
and investigated (for recent reviews see
papers\cite{fischer,weaver}). In the face-centered-cubic
(FCC) structure of pristine C$_{60}$ the molecules are
bonded via
van der Waals interaction. In this structure the tetrahedral
and octahedral interstitial spaces are sufficiently large to
accommodate alkali atoms without significant distortion of
the lattice. The intercalated alkali atoms transfer their
$s$ electron to C$_{60}$ molecules, consequently, the
lattice is contracted by the electrostatic interaction for
small atoms (e.g. Na).

Because of the charge transfer the electrostatic energy
plays an
important role in the formation of different ordered
structures, as well as in the distortion of the host lattice
for
large alkali concentration ($x>3$). Fleming {\it et al.}
\cite{fleming} have determined the electrostatic energies
for Rb$_x$C$_{60}$ compounds assuming point charges for the
Rb$^+$ ions, and the contribution of the C$_{60}^{x-}$ ion
is
expressed as the minimum Coulomb repulsion of $x$ point
charges on a sphere of radius $R$, with $R$ taken to be the
C$_{60}$ nucleon radius of 3.5 \AA. In the calculation of
electrostatic energies Rabe {\it et al.} \cite{rabe} have
assumed a uniformly charged spherical shell with the same
radius for the C$_{60}^{x-}$ ion.

In the present work we introduce a simple lattice-gas model
to study the ordering phenomena in alkali fullerides. In
this model the interstitial sites may be empty or occupied
by one type of alkali ion. In agreement with the large size
of fullerenes, here we assume that an A$^+$ alkali ion is
completely screened by uniformly distributing its
transferred charge on the first neighbor C$_{60}$ molecules.
For example, the alkali ion residing at a tetrahedral
(octahedral) site is surrounded by four (six) C$_{60}$
molecules with charges of $-e/4$ ($-e/6$). In other words,
each intercalated alkali atom transfers charges to its
neighboring C$_{60}$ molecules independently of the position
of the remaining A$^+$.
This simplification makes the derivation of an effective
pair interaction possible, which is evidently a
short range one. The difference between the sizes of
tetrahedral and octahedral voids is considered by
introducing extra site energy related to the van der Waals
interactions. This site energy parameter
depends on the radius of the intercalated particle,
consequently it is characteristic of the type of alkali
atom. In the knowledge of the effective interactions (and
site energy) characterized by a few parameters we are able
to study the ordering within the framework of the
lattice-gas (Ising) formalism. Accepting the spherical shell
charge
distribution on C$_{60}$ as suggested by Rabe {\it et al.}
\cite{rabe} this model reproduces the ordered ground state
energies given by them. A simple mean-field analysis is
carried out to demonstrate the richness of phase diagrams
when varying two parameters (site energy and radius of the
spherical shell).

The above formalism can describe the ordered structures
observed in alkali fullerides with FCC structure. Namely, in
pristine C$_{60}$ all the interstitial sites are empty,
whereas in the superconducting compounds with nominal
composition
A$_3$C$_{60}$ both the octahedral and tetrahedral sites are
occupied by alkali ions.\cite{stephens,hebard,ross,holczer}
Only the octahedral
sites are occupied in AC$_{60}$ forming NaCl structure for
A=Rb and Cs.\cite{nacl} Filling of the tetrahedral sites
results in the CaF$_2$ structure observed by Rossiensky {\it
et al.} \cite{caf2} in the Na$_x$C$_{60}$ system. According
to our analysis the above phases are found to be stable if
one takes the total energy of charged C$_{60}$ molecule into
consideration as discussed later on.

The early experimental investigations of the alkali
intercalated fullerides are summarized by Zhu {\it et
al.}\cite{zhu} in a provisional phase diagram.
Recently, Poirier and Weaver \cite{poirier} and Kuzmany {\it
et al.}\cite{kuzmany} proposed a phase
diagram for the K$_x$C$_{60}$ system showing a eutectoid
transformation from the homogeneous KC$_{60}$ state to the
composition of C$_{60}$ and K$_3$C$_{60}$
phases.\cite{poirier} Our calculation confirms the existence of
such a phase digram in a narrow range of parameters.

The above model explains adequately the formation of different
intercalated structures. Our purpose is not to argue that the present
description gives the correct lattice-gas model for a given A$_x$C$_{60}$
system but rather to emphasize some general features highlighted
by this approach because the model is adaptable for other
metal-C$_{60}$ systems. By this means the present model provides
a framework to classify these systems. The relevant features will
be illuminated by displaying a set of distinct phase diagrams.

In this paper we restrict ourselves to the systems
corresponding to A$_x$C$_{60}$ with rigid FCC structure. The
present description consider the C$_{60}$ molecule as a
spherical object, that is, the phenomena related to the
orientational ordering are neglected. Both the interaction
between two C$_{60}$ molecules and the electronic structure
are simplified therefore the present model can not describe
the chain formation discovered very recently by Chauvet {\it
et al}.\cite{janossy}

The lattice-gas model is derived in the subsequent section.
In Sec. \ref{sec:gs} we discuss the ground states suggested
by the model for different values of site energy.
A three-sublattice mean-field analysis is performed in Sec.
\ref{sec:mf} to demonstrate the variety of phase diagrams.
These results are confirmed by a simple (triangular) version
of the cluster-variation method (CVM) in Sec. \ref{sec:cvm}.
Finally we summarize our conclusions and discuss the
possible continuation of the present approach in Sec.
\ref{sec:conc}.

\section{THE MODEL}
\label{sec:model}

The electronic structure calculations for alkali fullerides
have indicated that the higher lying {\it s} bands of the
metal
are unoccupied and the electrons fill the empty $t_{1u}$
states of the C$_{60}$ molecule, so the alkali atoms are
ionized and their valence electrons are accepted by the
fullerene molecule forming a negative
ion.\cite{oshiyama,andersen} In such a situation the Coulomb
interaction has the most important contribution to the
energy of the system. Several authors have already studied
the
effect of the Madelung energy on the charge-transfer solids
\cite{hubbard}, binary \cite{magri,zunger} and pseudobinary
alloys \cite{schilfgaarde} and intercalated graphite
compounds \cite{metzger}.

  Statistical physics is able to give an adequate
description of the Ising systems with short range
interaction but a great deal arises in handling the infinite
range Coulomb interaction. An effective way of eliminating
this problem is to construct a renormalized short range
interaction.
For this purpose one can, for example, use the cluster
expansion of
the Madelung energy.\cite{zunger} In our treatment a well
converging series of pair interactions is derived in order
to avoid the difficulties arising from the long range order
interactions among the ions.

Calculations taking into account the polarizability of the fullerene
molecules in the host lattice \cite{pederson} suggested that the energy of two
charged C$_{60}$ molecules in the intercalated solid can be well
described as the interaction between two point charges in a dielectric
media characterized by $\varepsilon$.

According to that when calculating the Coulomb energy of the system, the
metal
atoms and C$_{60}$  molecules are treated as point charges.
The charge of the alkali atoms has been taken to be {\it e}
and the charge of the $C_{60}$ has been strongly related to
the occupation of the nearest neighbor interstitial sites.
The charge distribution may easily be described in the
lattice-gas formalism introducing the site occupation
variables $\eta_i$ which is 1 if the $i$-th interstitial
site is occupied and 0 if this site is empty. Within this
formalism the charge at site $i$ is evidently $q_i= e
\eta_i$ and the charge transferred to the $\alpha$-th
$C_{60}$ molecule can be written as:
 \begin{equation} \label{eq:charge}
  Q_{\alpha} =-e \sum_{j \in N_\alpha} c_j \eta_j
 \end{equation}
where $N_{\alpha}$ means the set of first neighbor
tetrahedral and octahedral interstitial sites around the
$\alpha$-th $C_{60}$ molecule and $c_j$ denotes the portion
of charge transferred from the alkali atom residing at the
$j$-th site to each nearest neighbor C$_{60}$ molecule. More
precisely, $c_j=1/6$ for octahedral and 1/4 for tetrahedral
sites. Such a choice provides that each alkali ion is
completely neutralized by its first neighbor C$_{60}$
molecules. The model assumes that the transferred charges
accompanies the alkali ions when hopping from site to site.
This charge distribution evidently satisfies the requirement
of the neutrality of the crystal.

The energy of the system can be given in the following form:
\begin{eqnarray}
E =&& {1\over 2\varepsilon} \sum_{i\neq j} {q_iq_j \over r_{ij}} +
{1\over 2\varepsilon} \sum_{\alpha \neq \beta} {Q_\alpha Q_\beta \over
r_{\alpha \beta}}  \nonumber \\
&+& \sum_{i,\alpha} {q_iQ_\alpha \over
\varepsilon r_{i\alpha}} + \sum_\alpha E_{int}(Q_\alpha)
\end{eqnarray}
where  $E_{int}(Q_\alpha)$ is the intramolecular energy, the Greek and Latin
indices run over the $C_{60}$
sites and interstitial sites of the FCC lattice
respectively. Substituting the charges from
(\ref{eq:charge}) the Coulomb energy can be expressed as:
\begin{eqnarray}
 E_C &=& {e^2\over 2\varepsilon} \sum_{i \neq j}{1\over r_{ij}} \eta_i
\eta_j  - {e^2\over \varepsilon}\sum_{i,\alpha} \sum_{j\in N_\alpha}{c_j \over
r_{i \alpha}} \eta_i \eta_j \nonumber \\
&& + {e^2\over 2\varepsilon}\sum_{\alpha\neq \beta} \sum_{j\in
N_\alpha} \sum_{k\in N_\beta} {c_j c_k \over r_{\alpha
\beta}} \eta_j \eta_k \; .
\end{eqnarray}

On exchanging the order of summations the energy can be
separated into the contributions of onsite and pair
interactions, namely
\begin{equation}
E_C = \sum_i \epsilon_i^{\prime} \eta_i + {1 \over 2}
\sum_{ij} V_{ij}^{\prime} \eta_i \eta_j
\label{eq:Ecoulomb}
\end{equation}
where
\begin{eqnarray}
\epsilon_i^{\prime} &=& {e^2 \over 2\varepsilon} \sum_{
(\alpha,\beta) \in N_i,          \alpha \neq \beta}
{c_i^2\over r_{\alpha\beta}}
- {e^2\over \varepsilon} \sum_{\alpha \in N_i} {c_j \over r_{i\alpha}}
\label{eq:Esite1} \;, \\
V_{ij}^{\prime} &=& {e^2\over \varepsilon r_{ij}} + {e^2\over \varepsilon}
\sum_{\alpha \in
N_i} \sum_{\beta\in N_j} c_i c_j { (1-\delta_{\alpha\beta})
\over r_{\alpha\beta}} \nonumber \\
&& - {e^2\over \varepsilon} \sum_{\alpha \in N_j}{c_j \over r_{i\alpha}}-
{e^2\over \varepsilon} \sum_{\alpha \in N_i}{c_i\over r_{j\alpha}}\;,
\label{eq:V1}
\end{eqnarray}
and $N_i$ denotes the set of first neighbor $C_{60}$
positions around interstitial site $i$.

The magnitude of $V_{ij}^{\prime}$ decreases rapidly with
the distance between sites $i$ and $j$.
One can realize in Eq.~\ref{eq:V1} the Coulomb interaction between two
cluster formed by the $i$--th and $j$--th alkali ion as the central site
and their nearest neighbor C$_{60}$ molecules. Since the clusters are
neutral and the systems have cubic symmetry the monopole, dipole and the
quadropole terms vanish. So
 there are only higher order multipole interactions between two sites.
The good convergence properties of the pair interaction is
a consequence of our basic assumptions, viz. the
nearest neighbor shell of the C$_{60}$ molecules completely
screens out the alkali ion charge.

In order to evaluate the intramolecular energy
a series of UHF-MNDO \cite{dewar} calculations
have been performed for up to a sixfold ionized single
C$_{60}^{x-}$ molecule
where the characteristic bond lengths of the system were kept constant
$r_1 = 1.400$ \AA and $r_2 = 1.440$ \AA\ .

\begin{figure}
\centerline{\epsfxsize=8cm
	    \epsfbox{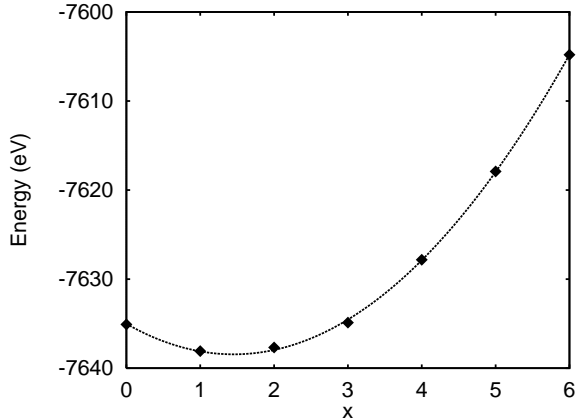}
	    \vspace*{2mm}}
\caption{Total energy of C$_{60}^{x-}$ ion. Diamonds are the
calculated values, dashed line represents the fitted
parabola.}
\label{fig:toten}
\end{figure}

The total energy of the charged molecule as a function of
the number of excess electrons can be fitted quite well with
a parabola (see Fig.~\ref{fig:toten}) given by:
\begin{equation}
   E_{tot} = A + Bx + C x^2
\label{eq:Ec60}
\end{equation}
where $x$ is the number of extra electrons on the molecule
and $A=-7635.02$\,eV; $B=-4.71926$\,eV; $C=1.6262$\,eV.
Such an almost perfect parabolic behavior does not hold for other
fullerenes with lower symmetry.

The quadratic energy dependence of the isolated molecule on the excess
charges allows us to introduce an effective radius for the C$_{60}$
molecule having a value of $R = 0.312 a$ where the lattice constant $a$
is given by experiment\cite{stephens}.
{}From previous calculations \cite{pederson,coulon,yamaguchi}
 $R \approx 4.8$ \AA \ can be obtained.
The differences between the present and the cited calculations are
due to the different choice of the equilibrium bond lengths.
Since the C$_{60}$ molecule is embedded in the crystal lattice the
energy of the single molecule is modified by the surrounding media.
Supposing  dielectric screening mechanism one can apply the
Onsager--Kirkwood theory of the electrostatic solvation \cite{kirkwood}
for estimating the effect of the lattice:
\begin{equation}
 E_{int} = E_0 - {1\over2} \left ( \left (1 - {1\over\varepsilon}\right )
 {Q^2\over r_0}
      + 2{\varepsilon-1\over 2\varepsilon +1}{\mu^2\over r_0^3} + ... \right )
\end{equation}
where $r_0$ is the radius of the cavity into which the molecule is
embedded. Since the $\mu$ dipole moment of the charged C$_{60}$ is zero
choosing the radius of the cavity to be equal the effective radius the
intramolecular energy can be written as:
\begin{equation}
 E_{int} = A + Bx + {e^2x^2\over 2\varepsilon R}
\end{equation}

 The first term gives a constant contribution to the total energy
and the linear term shifts the onsite energies. Both effects
may be transformed out by rescaling energy.
The spherical shell model proposed by Rabe {\it et al.} \cite{rabe}
with the original radius of the bucky--ball leads essentially different
ordering properties.
The quadratic
term gives a $V_{ij}^{\prime \prime}$ contribution to the
pair interaction between those alkali ions which transfer
some charge to the same C$_{60}$ molecule, i.e.:
\begin{equation}
V_{ij}^{\prime \prime} = {2\over\varepsilon} \sum_{\alpha} c_i c_j C
\label{eq:V2}
\end{equation}
where $\alpha \in N_i \cap N_j$. Obviously, this
contribution vanishes when $r_{ij}$ exceeds a threshold
value. Furthermore, the energy of C$_{60}^{x-}$ increases
the site energies too with
\begin{equation}
\epsilon_i^{\prime \prime}={1\over\varepsilon}\sum_{\alpha \in N_i}c_i^2 C
\;.
\label{eq:Esite2}
\end{equation}

The contribution of the van der Waals interaction to the
site energy  is distinct for the tetrahedral and octahedral
sites and depends on the size of the alkali ion. To
characterize
this contribution we introduce an additional site energy,
i.e.:
\begin{equation}
\epsilon_i=\left\{
\begin{array}{l}
\epsilon_o \;\;\;\;\hskip 8 mm \mbox{for octahedral
sites,}\\
\epsilon_t+\delta E_t \;\;\mbox{for tetrahedral sites}
\end{array}\right.
\label{eq:Esite}
\end{equation}
where $\delta E_t$ is considered as the difference between
the
two types of sites and is expected to be positive for large
alkali ions. In this expression the sum of
$\epsilon_i^{\prime}$ and $\epsilon_i^{\prime \prime}$
is indicated by $\epsilon_o$ and $\epsilon_t$ for the
octahedral and tetrahedral sites, respectively. The
numerical
calculations
give $\epsilon_o=-1.17817 e^2/{\varepsilon a}$ and $\epsilon_t=-
1.37843 e^2/{\varepsilon a}$ (where $e^2/a \approx 1$\, eV), i.e. the
tetrahedral sites are primarily occupied for $\delta E_t=0$.

We are now in a position to express the total energy of the
system within the lattice-gas formalism. The following
Hamiltonian defines the energy for any configuration of
alkali ions described by the variables $\eta_i$:
\begin{equation}
H = \sum_i \epsilon_i \eta_i + {1 \over 2} \sum_{ij}
V_{ij} \eta_i \eta_j \;,
\label{eq:Hamiltonian}
\end{equation}
where $V_{ij}=V_{ij}^{\prime}+V_{ij}^{\prime \prime}$ and
$\epsilon_i$, $V_{ij}^{\prime}$, $V_{ij}^{\prime
\prime}$ are defined above by Eqs. (\ref{eq:Esite}),
(\ref{eq:V1}) and (\ref{eq:V2}).

\begin{figure}
\centerline{\epsfxsize=8cm
	    \epsfbox{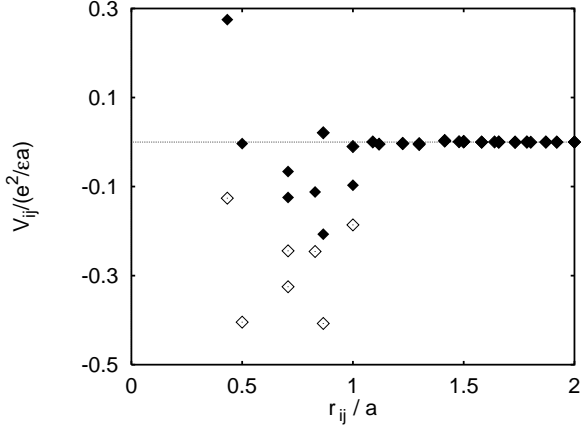}
	    \vspace*{2mm}}
\caption{Effective pair interaction as a function of ion-ion
distance. Open diamonds represent the pure coulombic
interaction ($V_{ij}^{\prime}$),  closed diamonds refer to
full interaction ($V_{ij}$) including the energy of charged
C$_{60}$ molecule.}
\label{fig:Vij}
\end{figure}

The effective pair interactions $V_{ij}$ vs. $r_{ij}/a$ are
shown in Fig.~\ref{fig:Vij}. $V_{ij}$ has different values
for the same distances; in other words, the pair interaction
depends on the orientation of the pair for some values of
$r_{ij}$. Figure \ref{fig:Vij} demonstrates clearly that
$V_{ij}$ becomes extremely weak if $r_{ij}>a$\,. The
attractive contributions of the effective pair interaction
(for $a/2<r_{ij} \leq a$) indicates the stability of alkali
intercalated fullerides at large alkali concentrations.

As a comparison $V_{ij}^{\prime}$ is also plotted in
Fig.~\ref{fig:Vij}. The striking difference between $V_{ij}$
and $V_{ij}^{\prime}$ indicates the important role of the
electronic energy of C$_{60}^{x-}$ ions as will be
investigated in detail in the subsequent sections.

\section{GROUND STATES}
\label{sec:gs}

\begin{figure}
\centerline{\epsfxsize=8cm
	    \epsfbox{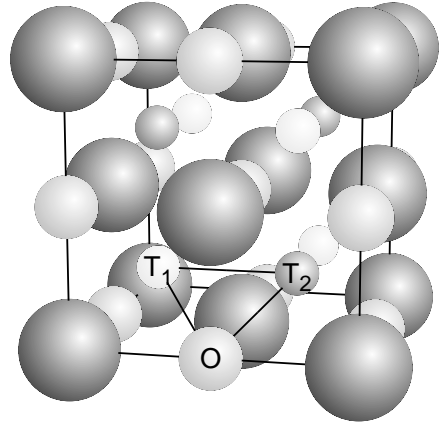}
	    \vspace*{2mm}}
\caption{Arrangement of octahedral (medium size spheres) and the
two tetrahedral (dark and light small spheres) interstitial sites in
an FCC lattice formed by C$_{60}$ molecules (large spheres).
The triangle represents the basic cluster of our CVM
approximation.
}
\label{fig:fcc}
\end{figure}

The electrostatic energies of some ordered structures have
already been investigated.\cite{fleming,rabe} Now we
concentrate on analysing such FCC structures which may be
described by
introducing three sublattices as demonstrated in
Fig.~\ref{fig:fcc}. The sites of each sublattice
form a FCC structure equivalent to the cage lattice. For
later convenience the sublattices are labelled 0
for octahedral and 1 or 2 for tetrahedral sites. In the
three-sublattice formalism the states are characterized by a
vector consisting of sublattice occupations. For example,
the state $(\sigma_0, \sigma_1, \sigma_2)$ denotes an alkali
distribution where the sites of sublattice $\nu$ are
occupied with a probability $\sigma_{\nu}$.
In general $0\leq \sigma_{\nu} \leq 1$, however,
$\sigma_{\nu}=$0 or 1 for ordered structures. These ordered
states are well known in crystallography. For example,
the states (1,0,0), (0,1,1) and (1,1,1) are equivalent to
the NaCl, CaF$_2$, and BiF$_3$ structures.

\begin{figure}
\centerline{\epsfxsize=8cm
	    \epsfbox{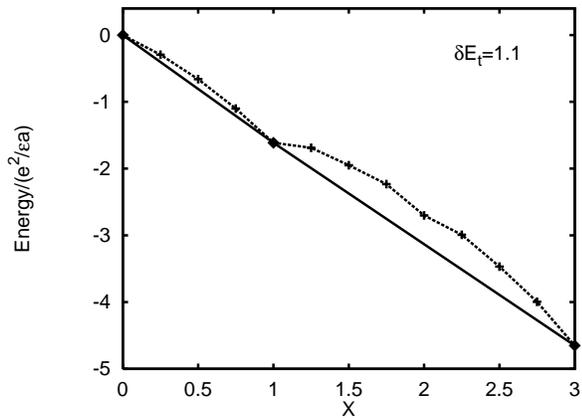}
	    \vspace*{2mm}}
\caption{Smallest electrostatic energies of the different
ordered
structures of A$_x$C$_{60}$ at fixed concentrations for
$\delta
E_t=1.1 e^2/\varepsilon a$. The solid line joins the stable phases.
}
\label{fig:gsen}
\end{figure}

One can easily see that according to the present model the
C$_{60}$ molecules are uniformly charged for the ordered
states defined above. Consequently, the electrostatic
energies are equivalent to the Madelung energies determined
previously.\cite{fleming,rabe} In
agreement with the expectation the energies of ordered
states are equivalent to those predicted by Rabe {\it et
al.}\cite{rabe} when choosing $R=3.5$\,\AA . In the above
model, however, we have chosen larger $R$ as detailed in the
previous section. This correction modifies the total energy
of the ordered states as plotted in Fig.~\ref{fig:gsen} for
$\delta E_t=1.1$\,. As a result the present model suggests
three stable ground states [(0,0,0), (1,0,0) and (1,1,1)],
which are connected by solid lines in the figure. That is,
the system with a nominal alkali content $x=1$ has
(1,0,0) phase in contrast to the former
model predicting a stable (0,1,0) phase.\cite{rabe}

\begin{figure}
\centerline{\epsfxsize=8cm
	    \epsfbox{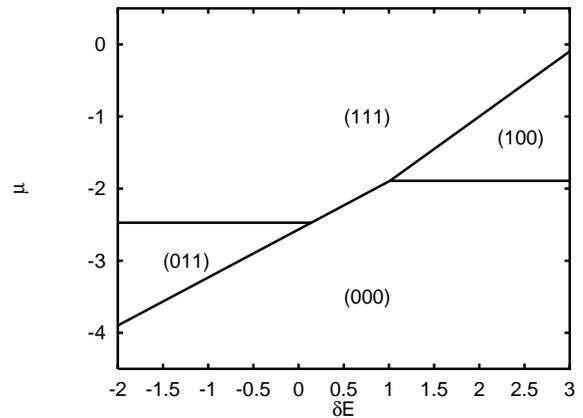}
	    \vspace*{2mm}}
\caption{Ground states of A$_x$C$_{60}$ as a function of
chemical potential $\mu$ and $\delta E_t$.}
\label{fig:gsmap}
\end{figure}

In order to study the stability of the ground states we
introduce the
chemical potential $\mu$ which controls the alkali
content ($x$) in the system. The stable state as a function
of $\mu$ is determined by the minimum of Gibbs potential per
C$_{60}$ defined as
\begin{equation}
{\cal G}_0={\cal H}_0-\mu x
\end{equation}
where the index $0$ refers to zero temperature and ${\cal
H}_0$ denotes the total energy per C$_{60}$. A simple
numerical calculation gives the stable state as a function
of $\mu$ and $\delta E_t$. The results are summarized in
Fig.~\ref{fig:gsmap}. This map shows that the alkali content
increases with the chemical potential for a fixed $\delta
E_t$. The energy relations are similar to those plotted in
Fig.~\ref{fig:gsen} if $\delta E_t > 1.0106 e^2/\varepsilon a$. There is
a region of $\delta E_t$ when the (0,0,0) state transforms
directly to (1,1,1) for increasing $\mu$ as is suggested
theoretically by Fleming {\it et al}.\cite{fleming} If
$\delta E_t<0.1406 e^2/\varepsilon a$ then the (0,1,1) state is stable.
This behavior is a consequence of the fact that $\delta E_t$
does not give an energy contribution to the states (0,0,0)
and
(1,0,0) however, it shifts the energy of the states (0,1,1)
and (1,1,1) by the same value.

We have evaluated the energies of several ordered states
related to a different sublattice division of the
tetrahedral sites. In order to determine the ground states
of the systems both the octahedral and tetrahedral (FCC)
lattices have been decomposed into four simple cubic
lattices and the energies of all the possible ordered
structures on the 12 sublattices have been calculated.

In comparison with previous results we obtained higher
energies for all the states
distinguishable from the former ones. This investigation
confirms the above sublattice division thereby providing a
framework for later analysis.

It is mentioned here that the ground state energies
are independent of the screening mechanism. The role of this
mechanism becomes important when studying the sublattice
occupations as a function of temperature.

\section{MEAN-FIELD APPROXIMATION}
\label{sec:mf}

In this section the thermodynamic properties of the
lattice-gas model defined by the Hamiltonian
(\ref{eq:Hamiltonian})
are investigated by using a three-sublattice mean-field
approximation. It is assumed that the interstitial points
are occupied by an alkali ion with the same probability
$\sigma_{\nu}$ within the sublattice $\nu = 0, 1, 2$.

By means of the mean-field approximation the energy per
C$_{60}$
molecule may easily be expressed by the variables
$\sigma_{\nu}$ as
\begin{eqnarray}
{\cal H}&=&\sum_{\nu} \epsilon_{\nu} \sigma_{\nu} + {1
\over 2} \sum_{\nu, \tau} J_{\nu \tau}\sigma_{\nu}
\sigma_{\tau}
\label{eq:Hmf}
\end{eqnarray}
where the $J_{\nu \tau}$ coupling constants contain all the
interactions between an alkali ion in sublattice $\nu$ and
those ions residing in sublattice $\tau$, i.e.,
\begin{equation}
J_{\nu \tau}= \sum_{j \in S_{\nu}} V_{ij} \;,\; \;\;\; (i
\in S_{\tau})\;.
\label{eq:J}
\end{equation}
where $S_{\nu}$ denotes the set of sites belonging to
sublattice $\nu$. The tensor of coupling constants is
symmetric, that is, $J_{\nu \tau}=J_{\tau \nu}$.
Furthermore, $J_{01}=J_{02}$ and $J_{11}=J_{22}$ because of
the equivalence of the tetrahedral sublattices.
Consequently, we have only four different coupling constants
characteristic of the model in mean-field approximation.
Their values may be determined by straightforward numerical
calculation which gives
\begin{eqnarray}
J_{00}&=& -1.42879 {e^2 \over \varepsilon a} \;, \nonumber \\
J_{11}&=& -1.60386 {e^2 \over \varepsilon a} \;, \nonumber \\
J_{01}&=& -0.29000 {e^2 \over \varepsilon a} \;, \nonumber \\
J_{12}&=& -0.86560 {e^2 \over \varepsilon a}
\label{eq:Jnum}
\end{eqnarray}
if $R/a=0.312$. Notice that all the $J_{\nu \eta}$ are
negative, i.e., the attractive forces rule over the short
range repulsion (see Fig.~\ref{fig:Vij}).

Here it is worth mentioning that Eq.~(\ref{eq:J}) permits
some modification of pair interactions $V_{ij}$ leaving the
mean-field parameters (as well as the ground state energy)
unchanged. For example, one can remove the interactions for
large distances adding suitable corrections to the short
range terms. In some sense the above screening mechanism may
be considered as an example of such a process. This
simplification may be very useful for Monte Carlo
simulations. As a limit case, one can introduce a model
characterized by the first four $V_{ij}$ terms with a
suitable choice of their strength.

It is evident that the mean-field Hamiltonian (\ref{eq:Hmf})
reproduces the energies of the ordered states discussed
above.
However, the coupling constants $J_{\nu \eta}$ cannot be
determined in the knowledge of the Madelung energies for all
the ordered structure because this Hamiltonian contains more
information, viz. it gives the potential energies for the
empty sites too.

The thermodynamic properties of the system are determined by
the minimization of Gibbs potential
\begin{equation}
{\cal G}={\cal H}-T {\cal S} - \mu \sum_{\nu} \sigma_{\nu}
\label{eq:G}
\end{equation}
with respect to $\sigma_{\nu}$ for fixed temperature T and
chemical potential $\mu$. In the above expression the
configurational entropy per C$_{60}$ is given as
\begin{equation}
{\cal S}=-k_B \sum_{\nu}[\sigma_{\nu}\log \sigma_{\nu}+ (1-
\sigma_{\nu}) \log (1-\sigma_{\nu})]
\label{eq:S}
\end{equation}
where $k_B$ is the Boltzmann constant.

The results of the numerical calculations are summarized in
temperature-composition ($T-x$) phase diagrams for different
values of $\delta E_t$. The alkali concentration is related
to sublattice occupations ($x=\sum \sigma_{\nu}$) determined
directly by the numerical procedure.

\begin{figure}
\centerline{\epsfxsize=8cm
	    \epsfbox{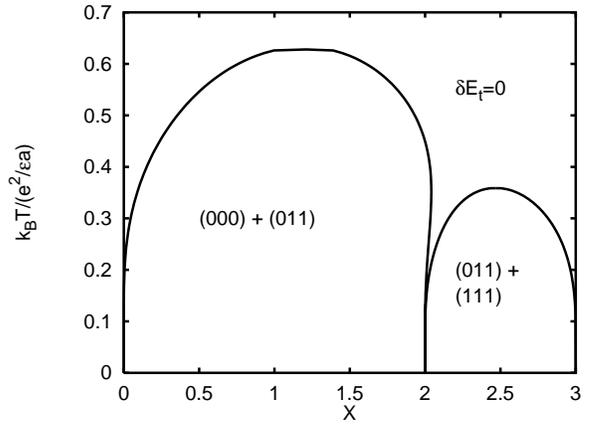}
	    \vspace*{2mm}}
\caption{Phase diagram for  $\delta E_t =0$ in mean-field
approximation.}
\label{fig:pd000mf}
\end{figure}

In agreement with the ground state analysis,
Fig.~\ref{fig:pd000mf} illustrates that only the (0,0,0),
(0,1,1) and (1,1,1) phases are stable at low temperatures
for $\delta E_t=0$. For intermediate compositions  the
system segregates into two phases as indicated in the
figure. It is emphasized that in the high temperature phase
the tetrahedral and octahedral sites are occupied with
different probabilities. More precisely, the tetrahedral
sites are preferred to the octahedral ones for $\delta
E_t=0$. For example, if the temperature is decreased for a
fixed
concentration $x=2$ the present model suggests a continuous
ordering process characteristic of the CaF$_2$-type
superionic conductors.\cite{sic}

\begin{figure}
\centerline{\epsfxsize=8cm
	    \epsfbox{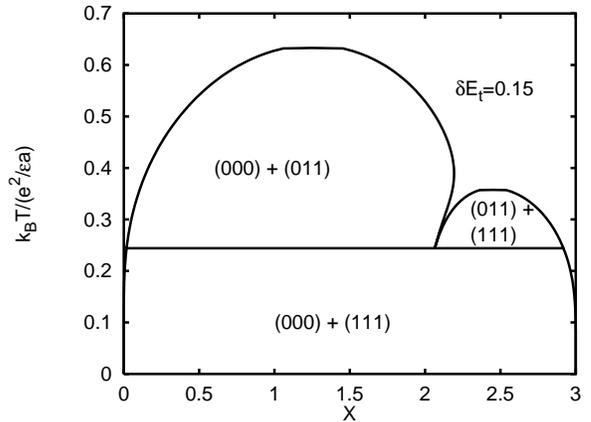}
	    \vspace*{2mm}}
\caption{Phase diagram in mean-field approximation for
$\delta E_t =0.15 e^2/\varepsilon a$ exhibits a eutectoid phase
transition.}
\label{fig:pd015mf}
\end{figure}

The (0,1,1) structure becomes unstable at $T=0$ when $\delta
E_t>0.1406 e^2/\varepsilon a $. Fig.~\ref{fig:pd015mf} shows a situation
when the homogeneous (0,1,1) phase may be observed in the
system but it decomposes into two states [(0,0,0)+(1,1,1)]
at low temperatures. A eutectoid transition appears at
$k_BT/(e^2/\varepsilon a)=0.2442$ for $\delta E_t=0.15 e^2/\varepsilon a$.
To demonstrate this eutectiod transition the value of $\delta
E_t$ is chosen to be close to the threshold value of its
stability at $T=0$.

The decomposition of the high temperature phase into the
phases (0,0,0) and (1,1,1) may be observed in a region of
$\delta E_t$ defined in Sec. \ref{sec:gs}\,. A typical phase
diagram that characterizes the phase separation at finite
temperatures will be shown in the subsequent section.

\begin{figure}
\centerline{\epsfxsize=8cm
	    \epsfbox{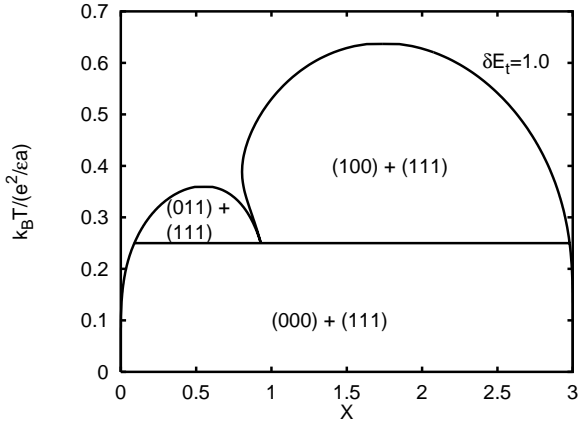}
	    \vspace*{2mm}}
\caption{Mean-field analysis of lattice-gas model suggests
eutectoid phase transition for  $\delta E_t =e^2/\varepsilon a$.}
\label{fig:pd100mf}
\end{figure}

As mentioned above the energy of both the (0,1,1) and
(1,1,1) phases increases with $\delta E_t$ meanwhile the
energy of the phases (0,0,0) and (1,0,0) remains unchanged,
and the (1,0,0) phase appears in the phase diagrams for
$\delta E_t>1.0106 e^2/\varepsilon a $. More precisely, close to this
stability threshold value one can observe a eutectoid
transition (see Fig.~\ref{fig:pd100mf}) which is analogous
to the situation discussed above. This phase diagram is
similar to those suggested by Poiriet and
Weaver\cite{poirier} and Kuzmany {\it et al.}\cite{kuzmany}
for the
K$_x$C$_{60}$ system for $x<3$.
With X-ray photoemission and Raman spectroscopy they
observed a reversible transformation from KC$_{60}$ to a
mixed phase of
C$_{60}$ and K$_3$C$_{60}$ at the eutectoid temperature
($T_{eut}=150 \pm 10\ ^{\circ}$C). For $\delta E_t=e^2/\varepsilon a$
the present approach predicts $k_BT_{eut}=0.250 e^2/\varepsilon a$.

A new type of phase diagram will be characteristic of the
system when the (1,0,0) phase becomes stable at zero
temperature. Instead of displaying now a typical plot, in
Sec.~\ref{sec:cvm} we compaire the phase diagrams suggested
by mean-field approximation and CVM for $\delta E_t=1.1
e^2/\varepsilon a$.

\begin{figure}
\centerline{\epsfxsize=8cm
	    \epsfbox{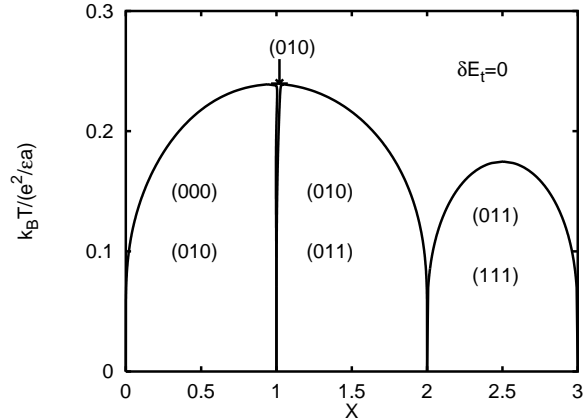}
	    \vspace*{2mm}}
\caption{Phase diagram for  $\delta E_t =0$ when the
electronic energy of C$_{60}^{x-}$ is approximated by a
spherical shell model with radius 3.5 \AA .}
\label{fig:pdrabe1}
\end{figure}

It is emphasized that in the above phase diagrams (including
those plotted in Sec.~\ref{sec:cvm}) both tetrahedral
sublattices are occupied with the same probability implying
the union of these sublattices. However, some modification
of the model parameters results in the appearance of the
(0,1,0) and (1,1,0) states.
We have also evaluated some phase diagrams accepting the
spherical shell model suggested by Rabe {\it et
al.}\cite{rabe} with a radius of the C$_{60}$ molecule. In
order
to demonstrate the difference between these choices we
display two phase diagrams corresponding to $R=3.5$ \AA . In
the first example the (0,1,0) state is stable at $T=0$ as
illustrated in Fig.~\ref{fig:pdrabe1}.
The additional feature observable in this plot is the
appearance of an order-disorder transition confined to
tetrahedral sites for $x=1$ meanwhile the octahedral sites
are practically empty. This process is analogous to an
``antiferromagnetic'' to ``paramagnetic'' transition, in
other words
the (0,1,0) state transforms into (0,1/2,1/2) at a critical
temperature ($k_B T_N=0.555 e^2/\varepsilon a$).

\begin{figure}
\centerline{\epsfxsize=8cm
	    \epsfbox{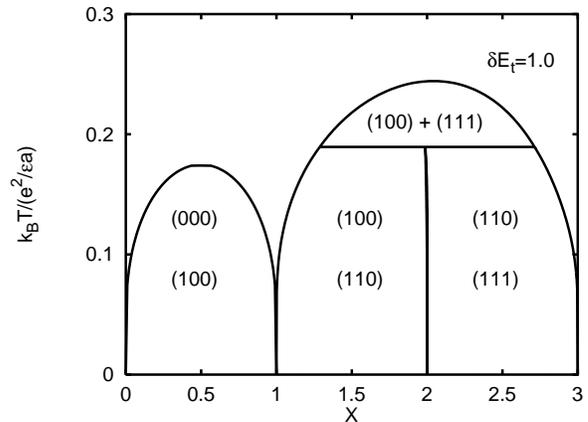}
	    \vspace*{2mm}}
\caption{Phase diagram for  $\delta E_t =e^2/\varepsilon a$ and $R= 3.5$
\AA .}
\label{fig:pdrabe2}
\end{figure}

The mean-field analysis suggests a very interesting phase
diagram for $\delta E_t=e^2/a$ (see Fig.~\ref{fig:pdrabe2}).
In this case, if the temperature for $x=2$ is decreased the
high
temperature phase decomposes into two phases whose alkali
concentrations are close to 1 and 3. At lower temperatures,
however, the (1,1,0) phase becomes stable in comparison to
the coexistence of states (1,0,0) and (1,1,1).

The richness of the predicted phase diagrams is surprising
because we have varied only two parameters. Rigorous
analysis of all the possible diagrams as a function of $R$
and $\delta E_t$, however, goes beyond the scope of the
present paper.

\section{CLUSTER-VARIATION METHOD}
\label{sec:cvm}

The cluster variation methods (CVM) are successfully used
for
deriving phase diagrams with sufficient
accuracy.\cite{revF} Two basic difficulties emerge when
adapting this technique to the present lattice-gas model.
On the one hand, the interstitial sites form a non-Bravais
lattice; on the other hand, instead of the usual first (and
second) neighbor interaction(s) there are a lot of
parameters
$V_{ij}$ characteristic of the interaction between sites $i$
and $j$. Here we suggest the application of a simple version
of CVM to increase the accuracy of the three-sublattice
mean-field approximation.

First we introduce a set of variables $t(\eta_0, \eta_1,
\eta_2)$ to denote the probability of a configuration
$(\eta_0, \eta_1, \eta_2)$ on three neighboring sites where
$\eta_{\nu}$ refers to the occupation of the site belonging
to sublattice $\nu$. The position of such a three-point
cluster is indicated by a triangle in Fig.~\ref{fig:fcc}.
The probability of a configuration on a part of this
triangle may be easily expressed by the values of $t$. For
example, the probabilities for pair configurations are given
as
\begin{eqnarray}
p_{01}(\eta_0,\eta_1) &=& \sum_{\eta_2} t(\eta_0, \eta_1,
\eta_2) \;, \nonumber \\
p_{02}(\eta_0,\eta_2) &=& \sum_{\eta_1} t(\eta_0, \eta_1,
\eta_2) \;, \\
p_{12}(\eta_1,\eta_2) &=& \sum_{\eta_0} t(\eta_0, \eta_1,
\eta_2) \;, \nonumber
\end{eqnarray}
and for single site configurations
\begin{eqnarray}
s_0(\eta_0) &=& \sum_{\eta1, \eta_2} t(\eta_0, \eta_1,
\eta_2) \;, \nonumber \\
s_1(\eta_1) &=& \sum_{\eta_0, \eta_2} t(\eta_0, \eta_1,
\eta_2) \;, \\
s_2(\eta_2) &=& \sum_{\eta_0, \eta_1} t(\eta_0, \eta_1,
\eta_2) \nonumber
\end{eqnarray}
whose values are directly related to the sublattice
occupations $\sigma_{\nu}$ introduced previously, namely
$s_{\nu}(1)=\sigma_{\nu}$ and $s_{\nu}(0)=1-\sigma_{\nu}$.

The contribution of the first and second neighbor
interactions to the system energy may be directly expressed
by the variables $p_{\nu \tau}(1,1)$ and only the remaining
terms are approximated on the basis of mean-field theory.
That is, now the energy per C$_{60}$ molecule is given as
\begin{eqnarray}
{\cal H}^{(CVM)}&=& 4 V(1) [p_{01}(1,1)+p_{02}(1,1)]+6 V(2)
p_{12}(1,1) \nonumber \\
&&+ {1 \over 2} \sum_{\nu, \tau}J_{\nu \tau}^{\prime}
s_{\nu}(1) s_{\tau}(1)+\sum_{\nu} \epsilon_{\nu}
s_{\nu}(1)
\label{eq:Hcvm}
\end{eqnarray}
where $V(1)=0.27459 e^2/\varepsilon a$ and $V(2)=-0.00386 e^2/\varepsilon a$
denotes
the interactions $V_{ij}$ at the shortest and second
shortest distances and $J_{\nu \eta}^{\prime}$ is equivalent
to $J_{\nu \eta}$ defined by Eq.~(\ref{eq:J}) except the
coupling constants should be reduced by the contributions of
$V(1)$ and $V(2)$,
\begin{eqnarray}
J_{01}^{\prime}&=&J_{01}-4 V(1) \;,\nonumber \\
&& \label{eq:Jcvm} \\
J_{12}^{\prime}&=&J_{12}-6 V(2) \;. \nonumber
\end{eqnarray}

Here we restrict ourselves to demonstrating how the entropy
may be derived on the analogy of
pair approximation developed by Bethe.\cite{bethe}
The non-Bravais lattice of interstitial sites may be built
up by repeating periodically the triangle of an octahedral
and two tetrahedral sites as prescribed by the primitive
cell of FCC structure. This construction makes it clear that
each triangle
is surrounded by twelve triangles with the same orientation.
According to the pair approximation the entropy is expressed
by the probability of pair configurations.
\cite{bethe,pairfcc} In the present situation the
probability of a configuration on a pair of triangles is
approximated as a product of $t(\eta_0, \eta_1,\eta_2)$
variables. The calculation leads to the following formula:
\begin{eqnarray}
{\cal S}^{CVM}&=& -6 k_B \sum_{\eta_0, \eta_1, \eta_2}
t(\eta_0, \eta_1, \eta_2) \log t(\eta_0, \eta_1, \eta_2)
\nonumber \\
&&+k_B \sum_{\nu < \tau} x_{\nu \eta} \sum_{\eta_{\nu},
\eta_{\tau}} p_{\nu \tau} (\eta_{\nu}, \eta_{\tau}) \log
p_{01} (\eta_{\nu}, \eta_{\tau}) \nonumber \\
&&+k_B \sum_{\nu} y_{\nu} \sum_{\eta_{\nu}} s_{\nu}
(\eta_{\nu}) \log s_{\nu} (\eta_{\nu})
\label{eq:Scvm}
\end{eqnarray}
where $x_{01}=x_{02}=y_1=y_2=2$ and $x_{12}=y_0=1$.

Substituting Eqs.~(\ref{eq:Hcvm}) and (\ref{eq:Scvm}) into
(\ref{eq:G}) one obtains the Gibbs potential as a function
of $t(\eta_0,\eta_1,\eta_2)$ variables. As is well known,
${\cal
G}$ reaches its minimum at the equilibrium values of these
variables. In fact we have seven independent variables
because the eighth configuration probability is determined
by the condition of normalization. The numerical treatment
is similar to those followed in mean-field approximation.
The only difference, that instead of the three
$\sigma_{\nu}$
variables now we have seven parameters to minimize the Gibbs
potential with respect to them.

\begin{figure}
\centerline{\epsfxsize=8cm
	    \epsfbox{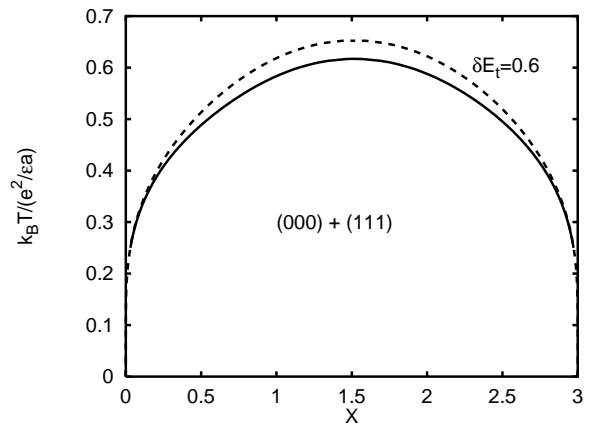}
	    \vspace*{2mm}}
\caption{Comparison of phase diagrams suggested by
mean-field approximation  (dashed line) and CVM (solid line)
for  $\delta E_t =0.6 e^2/\varepsilon a$.}
\label{fig:pd060}
\end{figure}

\begin{figure}
\centerline{\epsfxsize=8cm
	    \epsfbox{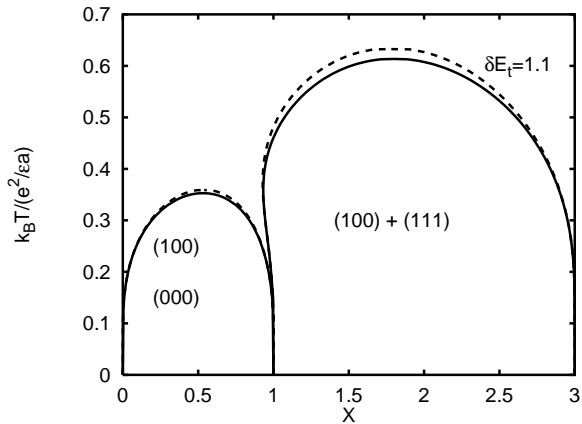}
	    \vspace*{2mm}}
\caption{Phase diagram for  $\delta E_t =1.1 e^2/\varepsilon a$
determined by mean-field approximation (dashed line) and CVM
(solid line).}
\label{fig:pd110}
\end{figure}

All the (mean-field) phase diagrams for $R/a=0.312$ have
been recalculated by using this technique. The comparisons
between the results of mean-field approximation and CVM are
demonstrated in Figs.~\ref{fig:pd060} and
\ref{fig:pd110}. These phase diagrams are typical in the
present model for the regions $0.1402< \delta E_t/ (e^2/a) <
1.0106$ and $\delta E_t > 1.0106 e^2/a$ respectively. In
both cases the CVM suggests lower transition temperatures in
comparison with mean-field results. In general we can say
that the difference is only a few percent for $R/a=0.312$;
however, it becomes larger for smaller $R$ when the short
range interactions are more repulsive.

In brief, this version of CVM confirms qualitatively the
results of mean-field approximation and leaves its main
conclusions unchanged.

\section{SUMMARY AND CONCLUSIONS}
\label{sec:conc}

We have developed a lattice-gas model for investigating the
formation of different alkali intercalated fullerides with
FCC structure. In this (Ising type) model the tetrahedral
and octahedral interstitial sites of the cage lattice are
empty or singly occupied by one type of alkali ion. It is
assumed that the Coulomb interaction between two alkali ions
is screened out by distributing their $s$ electrons
uniformly on their nearest neighbor C$_{60}$ molecules. This
plausible assumption results in a short range interaction
which is very convenient for the lattice-gas formalism. For
ordered structures this model reproduces the electrostatic
energies found by previous authors.\cite{fleming,rabe} At
the same time this approach allows us to study the effect of
electrostatic forces on the formation of different ordered
and disordered structures at finite temperatures.

Besides the electrostatic energy the model takes into
account the electronic energy of  C$_{60}^{x-}$ ions as well
as the van der Waals interaction between an alkali ion and a
C$_{60}$ molecule. A series of UHF-MNDO calculations shows
that the total energy of a charged C$_{60}$ molecule has a
term proportional to the square of its charge. This
nonlinear contribution mediates an interaction between
those alkali ions which transfer electrons to the same
C$_{60}$ molecule. This contribution is equivalent to the
electrostatic energy of a charged spherical shell with
radius $R$.
The extra site energy ($\delta E_t$) characteristic of the
van der Waals interaction at the smaller (tetrahedral)
interstitial voids makes a distinction between the different
types of alkali atoms. Without this term the tetrahedral
sites are preferred in the system. In agreement with
previous expectations\cite{weaver,caf2} this energy
contribution increases with the size of alkali atom and the
octahedral sites are primarily occupied when $\delta E_t$
exceeds a threshold value.

The thermodynamic properties of this model are investigated
by using a three-sublattice mean-field approximation. The
general features are illustrated via a series of phase
diagrams. These diagrams draw a picture of the effect of van
der Waals interaction and the electronic energy of the
C$_{60}^{x-}$
ion on the ordering process. These results are confirmed
qualitatively by CVM.

In the present approach the dielectric constant $\varepsilon$ is
considered as a parameter. A possible value ($\varepsilon =9$) may be
estimated by fitting the experimental Raman shift in the A$_6$C$_{60}$
fullerides.\cite{sanguinetti} Different values ($\varepsilon \approx 4$) are
suggested by Pederson and Quong \cite{pederson} from the polarizability
of C$_{60}$ molecules. Desregarding the above uncertainty the maximum
of the predicted ordering temperatures agree qualitatively with those
expected on the basis of recent experiments. There are more
reliable data on the phase separation temperature observed in the
K$_x$C$_{60}$ system. The present model can describe such type of
phase diagrams (see Fig.~\ref{fig:pd100mf} ) in a very narrow range
of $\delta E_t$ in
agreement with the fact that this feature has not been observed
for other fullerides.\cite{koller}
Furthermore, applying pressure in order to contract the lattice
the increasing $\delta E_t$ will couse the sharp decrease of the euthectic
temperature. Further increase of the pressure can even eliminate the
phase separation and possibly an orthorombic distortion of the cubic
structure will appear typical of the A$_1$C$_{60}$ systems at low temperature.

The ground state energy as a function of $\delta E_t$ and the series
of phase diagrams describe clearly that the A$_2$C$_{60}$ composition
can appear only for sufficiently low $\delta E_t$ values corresponding
to small atomic sizes as observed for A=Li and Na.

The generalization of the present description for alkaline earth
(or other) metals is straigthforward. If the intercalated atoms
have $z$ valence electrons we should substitute a new energy unit,
$(ze)^2/(\varepsilon a)$ for $e^2/(\varepsilon a)$ and find a suitable
$\delta E_t$ expressed in this unit. Due to the enhanced energy unit
the A$_2$C$_{60}$ composition can appear in a wider range of
$\delta E_t$. At the same time the stronger Coulomb forces contract
the lattice and modify the ratio $R/a$ which can lead to some
change in the phase diagrams.

The present approach is based on a simplified screening
mechanism. This hypothesis may be very fruitful for
investigating other intercalation compounds. The analysis of
alkali intercalated fullerides with body-centered-cubic
and -tetragonal structure is in progress.

\acknowledgements

This research was supported in part by the Hungarian
National Research Fund (OTKA) under Grants Nos. 2960 and
F014378.

\end{document}